# Experimental studies of THGEM in different Ar/CO$_2$ mixtures

He Zhan-ying(何占营)[1;2]   Zhou Jian-rong(周健荣)[2;3*]   Sun Zhi-jia(孙志嘉)[2;3]
Yang Gui-an(杨桂安)[2;3]   Xu Hong(许虹)[2;3]  Wang Yan-feng(王艳凤)[2;3]
Liu Qian(刘倩)[3;4]   LIU Hong-bang(刘宏邦)[3;4]   Chen Shi(陈石)[4]
Xie Yi-gang(谢一冈)[2;4]   Zheng Yang-heng(郑阳恒)[3;4]   Wang Xiao-dong(王晓冬)[1]
Zhang Xiao-dong(张小东)[1]   Hu Bi-tao(胡碧涛)[1]   Chen Yuan-Bo(陈元柏)[2;3]

[1] School of Nuclear Science and Technology, Lanzhou University, Lanzhou 730000, China;
[2] Institute of High Energy Physics, Chinese Academy of Sciences, Beijing 100049, China;
[3] State Key Laboratory of Particle Detection and Electronics，Beijing 100049，China;
[4] University of Chinese Academy of Sciences, Beijing 100049，China

**Abstract:**

In this paper, the performances of a type of the domestic THGEM (THick Gaseous Electron Multiplier) working in the Ar/CO$_2$ mixtures are reported in details. This kind of single THGEM can provide the gain range from 100 to 1000, which is very suitable for the application in the neutron detection. In order to study its basic characteristics as the references for the development of THGEM based neutron detector, the counting rate plateau, the energy resolution and the gain of the THGEM have been measured in the different Ar/CO$_2$ mixtures with the change of the electrical fields. For the Ar/CO$_2$(90%/10%) gas mixture, a wide counting rate plateau is got from 720V to 770V with the plateau slope of 2.4% / 100 V and the excellent energy resolution about 22% is obtained at the 5.9keV full energy peak of the $^{55}$Fe X-ray source.

## 1. Introduction

Neutrons are used to investigate the structure and dynamics of a material. Many efforts have recently been devoted to the development of the next generation of neutron facilities, which include SNS in USA, J- PARC in Japan, ISIS in UK, CSNS (China Spallation Neutron Source) in China and ESS in Europe [1]. The neutron detector is one of the key components of the neutron scattering instruments. With the international development of the new generation neutron source, the traditional neutron detector based on He-3 has not been able to satisfy very well the demand of the application of high flux especially. And also facing the global crisis of He-3

* Supported by National Natural Science Foundation of China (11127508 ,11175199)

*) E-mail: zhoujr@ihep.ac.cn

Phone:010-88236411

supply as well [2], the research on the new style of the neutron detector which can replace the He-3 based detection technology becomes extremely urgent.

As a good candidate, a boron coated GEM became the focus of attention recently [3], which was firstly designed by Martin Klein using CERN standard GEM in 2006[4]. It has the outstanding and excellent characteristics, such as high counting rate capability (>10MHz), good spatial resolution and timing properties, radiation resistance, flexible detector shape and readout patterns [5]. In 2011, IHEP and UCAS firstly developed successfully a kind of THGEM, manufactured economically by standard printed-circuit drilling and etching technology in China. Compared with the CERN standard GEM, THGEM has higher gain, sub-millimeter spatial resolution and the possibility of industrial production capability of large-area robust detectors, which is very suitable and adequate for the application of neutron detection.

In this paper, the THGEM is provided by Zheng Yang-heng group of UCAS. It's a thinner-THGEM with a thickness of 200μm, hole diameter of 200μm, pitch of 500μm and a very small rim of 5–10μm. The thinner-THGEM is made of FR4 glass epoxy substrate, with 20μm thick copper- cladding on both sides and has an active area of 50mm*50mm [6].In order to study its basic characteristics as the references for the development of this kind of domestic THGEM based neutron detector, the performances of the counting rate plateau, the energy resolution and the gain have been measured in the different $Ar/CO_2$ mixtures with the different high voltages. According to the experiments, the working conditions optimized of the THGEM have been obtained, which would be very helpful for the design of this kind of THGEM based neutron detector in future [7,8].

## 2. Experimental setup

Fig.1 shows a schematic view of the detector configuration consisting of cathode, anode and a single THGEM. The THGEM detector was operated in different $Ar/CO_2$ mixtures at a normal pressure and temperature. Measurements were carried out by using a $^{55}$Fe X ray source (activity 10mCi) which was positioned in such a way that a collimated beam (ϕ1 hole) of X-rays perpendicularly entered the upper drift region. The ionization electrons generated by the interaction of $^{55}$Fe 5.9keV X-rays with Ar atoms were amplified in avalanche mode in the THGEM holes [8], and then entered the induction region, where they were finally collected by the anode. All the three electrodes of HV were supplied by the WIENER MPOD mini-HV power and the signals were readout with an ORTEC 142IH preamplifier followed by an ORTEC 572A amplifier (shaping time t=2μs) and an ORTEC multi-channel analyzer (trump-usb-8k).

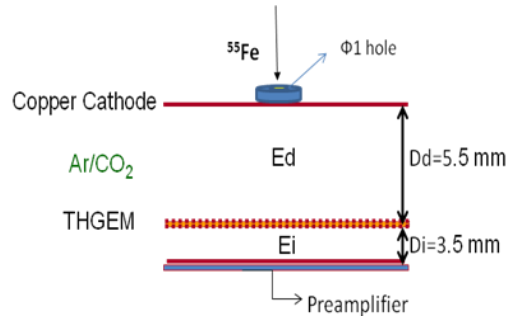

Fig.1 Schematic view of a single THGEM detector with drift region Dd=5.5mm and induction region Di=3.5mm.

## 3. Results and discussions

### 3.1 Plateau

In order to know the suitable working voltage of the detector in different $Ar/CO_2$ mixture, its counter plateau was measured in different Ed (drift field) and Ei (induction field) and the results are shown on Fig.2. During the experiment, the total flow of Ar and $CO_2$ gas is 50 SCCM to ensure the amount of effective working gas in the chamber. The counts were recorded in every one minute and the voltage of THGEM was increased by the increment of 5V until the spark discharge occurred. As the Fig.2 shows, it has a longer plateau in the $Ar/CO_2$ mixture ratio of 90%/10% and gets shorter with the increase of the proportion of $CO_2$. In certain extent, the counting rate was independent on the Ed and increasing with the Ei, so that the plateau was shifted left with the increasing of the Ei. For the gas mixture $Ar/CO_2$ (90%/10%), the induction field Ei of 2.0 kV/cm and the drift field Ed of 0.5kV/cm, the plateau range of this kind of THGEM was from 720V to770 V and its plateau slope was about 2.4% / 100 V, as shown in the top left of the Fig.2. This optimization would be helpful to know the working range of the THGEM and find the conditions with the lower HV.

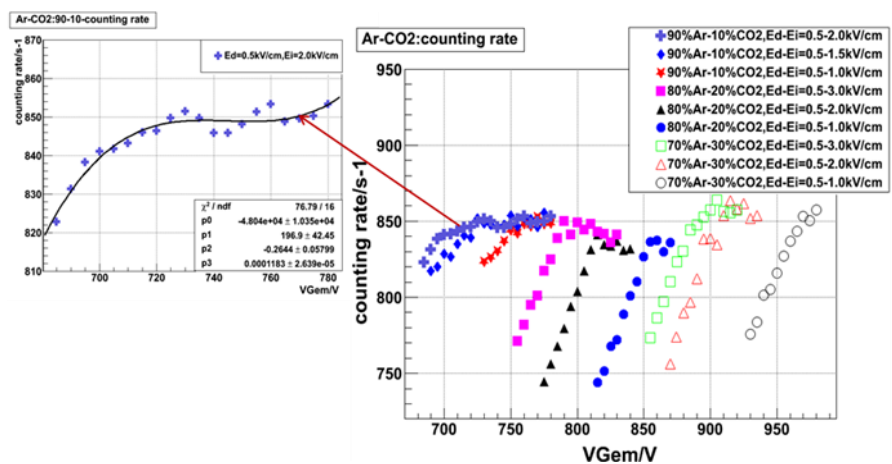

Fig.2 Counting rate plateaus of the THGEM detector in the three different kinds of gas mixture (90%Ar/10%$CO_2$、 80%Ar/20%$CO_2$ and 70%Ar/30%$CO_2$) with Ed=0.5kV/cm and Ei=1.0kV/cm、1.5kV/cm、2.0kV/cm and 3.0kV/cm.

## 3.2 Gain

By using the 5.9keV full energy peak of the $^{55}$Fe X-ray source, the effective gain was measured with the voltage of THGEM increased by the increment of 5V in different Ar/$CO_2$ mixture and different fields Ed and Ei. In certain extent, the stronger the induction field Ei is and the greater the voltage of the THGEM is, the gain is larger for the same gas mixture. As Fig.3 shows, the gain increases exponentially with the voltage of the THGEM for each kind of gas mixture and the single THGEM can provide a wide gain range from 100 to1000. The gain decreases with the increase of the proportion of $CO_2$ in the gas mixture.

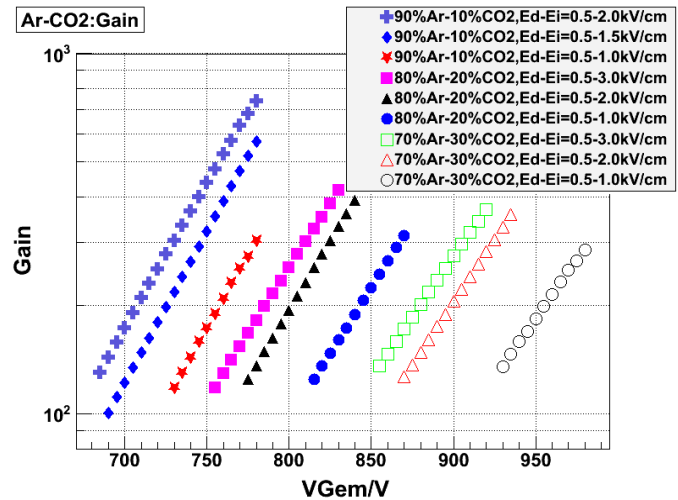

Fig.3 The experimental gain in the three different kinds
of gas mixture (90%Ar/10%$CO_2$、80%Ar/20%$CO_2$ and
70%Ar/30%$CO_2$) with Ed=0.5kV/cm and Ei=1.0kV/cm、
1.5kV/cm、2.0kV/cm and 3.0kV/cm.

As mentioned above, the influence of drift field Ed to the gain is very small in certain extent. Fig.4 (a) shows the gain as a function of Ed in different Ei in the Ar/$CO_2$ (90%/10%) gas mixture. In different Ed, the gain keeps nearly no change at the voltage 770V of THGEM for each Ei and increases with the Ei. As the same, Fig.4 (b) shows the gain change with Ed in different Ei in the Ar/$CO_2$(80%/20%) gas mixture and it shows the similar results as those in the Ar/$CO_2$ (90%/10%) gas mixture. For the Ar/$CO_2$ (70%/30%) gas mixture, the same regularity exists. Due to no common appropriate voltage of THGEM, the similar kind of this figure has not been presented.

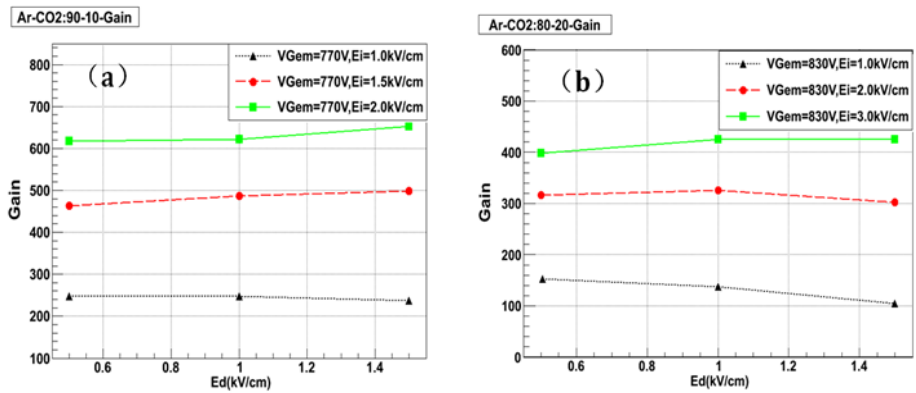

Fig.4 Gain vs Ed is in different Ei and in same voltage of the THGEM. (a) Gain changes with the Ed in an Ar/$CO_2$ (90%/10% ) gas mixture and the voltage of the THGEM VGem=770V.(b) Gain changes with the Ed in an Ar/$CO_2$ (80%/20%) gas mixture and the voltage of the THGEM VGem=830V.

### 3.3 Energy resolution

Energy resolution is one of the most important parameters related to the detector performance. By using the 5.9 keV full energy peak of the $^{55}$Fe X-ray source, the energy resolution was measured. The Fig.5 shows the results of three kinds of gas mixture with the voltage of THGEM increased by the increment of 5V in different field Ei. There are three regions with clear boundary related to the ratio of $CO_2$ in the gas mixture. As the ratio of $CO_2$ decreases, the working voltage of the THGEM will get lower and the energy resolution will get smaller as well. The much better energy resolution is obtained in the Ar/$CO_2$ (90%/10%) gas mixture. With the voltage of THGEM increasing, the energy resolution will also get better obviously. The induction field Ei has a bit affection on the energy resolution and the drift field has nearly no effect on the energy resolution from 0.5 to 3kV/cm (measured and not included in the figure). As a summary for the optimization, the best working points of this kind of THGEM are recommended as following: the Ar/$CO_2$ (90%/10%) gas mixture, the drift field 0.5kV/cm, the induction field 2kV/cm and the voltage range of THGEM from 720V-770V, which will give the better energy resolution smaller than 25%.

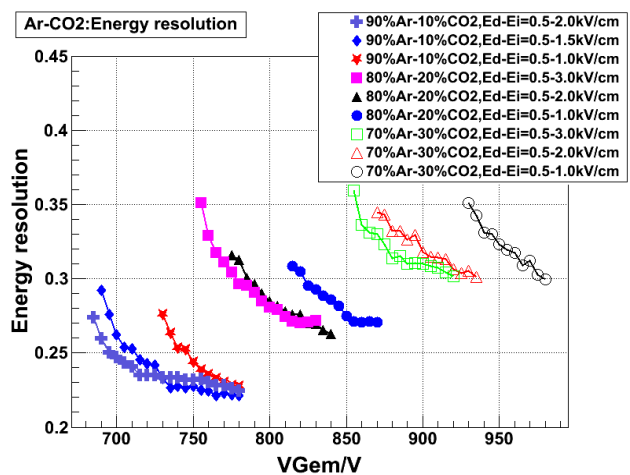

fig.5 The energy resolution in the three different kinds of gas mixture (90%Ar/10%$CO_2$、80%Ar/20%$CO_2$ and 70%Ar/30%$CO_2$) with Ed=0.5kV/cm and Ei=1.0 kV/cm、1.5 kV/cm、2.0 kV/cm and 3.0kV/cm.

Fig.6 shows a pulse height spectrum obtained with a $^{55}$Fe source in the Ar/$CO_2$ (90%/10%) gas mixture. For obtaining energy resolution, it's fitted with the Gaussian function. It indicates the energy resolution (FWHM) of the detector based on THGEM is about 22%. With such an energy resolution, the detector can entirely separate the 3keV of Ar escape peak from the $^{55}$Fe main X-ray peak located at 5.9keV.

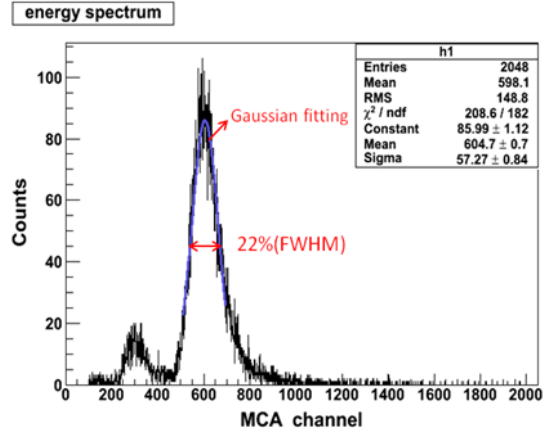

Fig.6 A pulse height spectrum in the Ar/$CO_2$ (90%/10%) gas mixture、Ed=0.5kV/cm、Ei=2kV/cm and the voltage of the THGEM VGem=770V with $^{55}$Fe source.

## 4. Conclusion

In this paper, the experimental performance of a kind of domestic THGEM working in the Ar/$CO_2$ mixtures are presented in details. The effective gain of single THGEM can reach about 1000 in the Ar/$CO_2$ mixture. As the ratio of $CO_2$ decreases from 30% to 10%, the working voltage of the THGEM will get lower, the plateau will get longer and the energy resolution will get much better as well. With the induction field Ei increasing from 1 to 3kV/cm, the performance of THGEM will be getting better and better. The drift field Ed (0.5 to 1.5kV/cm) has nearly no influence on the performance of THGEM. As a summary for the optimization, the best working points of this kind of THGEM are recommended as following: the Ar/$CO_2$ (90%/10%) gas mixture, the drift field 0.5kV/cm, the induction field 2kV/cm and the voltage range of THGEM from 720V-770V, which will give the better energy resolution smaller than 25%. According to the experiments, it would be very helpful for the design of this kind of THGEM based neutron detector in future.

## 5. Acknowledgements

We are grateful to the supports from National Natural Science Foundation of China (item: 11127508 and 11175199), China Spallation Neutron Source, the State Key Laboratory of Particle Detection and Electronics and the State Key Laboratory of neutron detection and fast electronics technology in Dongguan University of Technology.

## References

[1]B.Gebauer, Towards detectors for next generation spallation neutron sources, Nucl.Instr. and Meth. A,2004, 535: 65-78

[2] Dana A. Shea et al. The Helium-3 Shortage: Supply, Demand, and Options for Congress, Congressional Research Service, 2010

[3]H. Ohshita et al. Development of a neutron detector with a GEM, Nucl.Instr. and Meth. A, 2010,623:126-128

[4] C.Schmidt, M.Klein, The CASCADE Neutron Detector: A System for 2D Position

Sensitive Neutron Detection at Highest Intensities, Neutron News, 2006, 17(1):12-15

[5] M.Klein, C.Schmidt, CASCADE neutron detectors for highest count rates in combination with

ASIC/FPGA based readout electronics, Nucl.Instr. and Meth. A, 2011, 628:9-18

[6] H.B. Liu, Y.H.Zheng et al. The performance of thinner-THGEM, Nucl.Instr. and Meth. A ,2011,659:237–241

[7] Zhou JianRong et al. Neutron beam monitor based on a boron-coated GEM, Chinese Physics C,2011,35(7):668-674

[8] WANG Yan-Feng, SUN Zhi-Jia, ZHOU Jian-Rong et al. Simulation study on the boron-coated GEM neutron beam monitor, Chinese Physics, SCIENCE CHINA Physics,Mechanics & Astronomy,2013(received)

[9] A.Breskin, R.Alon et al.A concise review on THGEM detectors, Nucl.Instr. and Meth. A ，2009，598:107–111